\documentclass[12pt]{article}
\usepackage{a4wide}

\def\bm#1{\mbox{\boldmath{$#1$}}}

\def\ra{\rightarrow}

\def\Ra{\Rightarrow}

\def\p{\partial}

\def\d{\delta}

\def\l{\lambda}

\def\th{\theta}
\def\t{\tau}

\def\P{{\cal P}}

\def\S{{\cal S}}

\newcommand{\be}{\begin{equation}}
\newcommand{\ee}{\end{equation}}
\newcommand{\bea}{\begin{eqnarray}}
\newcommand{\eea}{\end{eqnarray}}

\begin{document}

\title{Stochastic formulation of the renormalization group:
supersymmetric structure and topology\\ of the space of couplings}

\author{Jos\'e Gaite\\
{\small
Instituto de Matem{\'a}ticas y F{\'\i}sica Fundamental,
CSIC, Serrano 123, 28006 Madrid, Spain}
}

\date{19 April 2004}

\maketitle

\abstract{The exact or Wilson renormalization group equations can be
formulated as a functional Fokker-Planck equation in the
infinite-dimensional configuration space of a field theory, suggesting
a stochastic process in the space of couplings.  Indeed, the ordinary
renormalization group differential equations can be supplemented with
noise, making them into stochastic Langevin equations.  Furthermore,
if the renormalization group is a gradient flow, the space of
couplings can be endowed with a supersymmetric structure {\em a la}
Parisi-Sourlas.  The formulation of the renormalization group as
supersymmetric quantum mechanics is useful for analysing the topology
of the space of couplings by means of Morse theory. We present simple
examples with one or two couplings.}

\section{Introduction}

The concept of the renormalization group arose in quantum
electrodynamics and was soon applied to other quantum field theories
and later to critical phemomena. With the application of the
renormalization group (RG) to several couplings, it became clear that
it could have interesting features as a system of autonomous ordinary
differential equations and, in particular, that the topology of the RG
trajectories should play a crucial role \cite{WilKog}. The simplest
topologies correspond to trajectories that follow the gradient of some
potential. This gradient RG flow hypothesis was discussed in
Ref.~\cite{WallZia}. With the generalization of this hypothesis to the
existence of an irreversible RG function, after Zamolodchikov
$c$-theorem in two dimensions \cite{Zamo}, it has been the subject of
numerous papers (as a representative sample, see \cite{All,IOCon,I1}).

However, the study of the topology of the space of couplings of a
field theory is still in its infancy. Even under the assumption of
gradient RG flow (or irreversible RG function) very few general
results exist. In two dimensions the problem has received more
attention, because of the powerful methods provided by conformal
symmetry and the connection with string theory. A particularly
interesting development in this regard is the relation with
supersymmetric quantum mechanics (SUSY QM) and Morse theory, two
concepts which were connected in Witten's seminal papers \cite{Wit},
independently of the RG.

Das, Mandal and Wadia proposed the connection of two-dimensional RG
equations with stochastic quantization and supersymmetric quantum
mechanics in the context of string theory \cite{DasManWa}. The
motivation was that two-dimensional quantum field theories are the
basis of first quantized (``classical") string theory and the field
equations are given by conformal invariance, that is, by the vanishing
of the $\beta$-functions corresponding to low-energy fields, which
play the role of couplings. Therefore, the interpolation between RG
fixed points, given by the RG flow, represents the transition between
string theory solutions, and a potential for the flow is also a
low-energy string potential. In this context, it is natural to
introduce supersymmetry in the space of couplings, now low-energy
fields.  The underlying supersymmetry of stochastic quantization had
been discovered earlier by Parisi and Sourlas \cite{ParSour} (for a
systematic treatment, see \cite{ZJ}). In the string theory context, it
is natural to assume that the fields have a stochastic character and,
in fact, this character corresponds to quantized string field theory,
that is, to {\em second-quantized} string theory.

A different point of view was adopted by Vafa \cite{Vafa}, regarding
the topology of the space of two-dimensional quantum field theories as
given by Zamolodchikov's $c$-function when considered as a Morse
function.

We adopt here a more general standpoint: the field theories need not
be two-dimensional and, hence, need not have relation with string
theory. Supersymmetry in the space of couplings is just a convenient
mathematical structure to study the topological structure of this
space, following the spirit of Witten's paper ``Supersymmetry and
Morse Theory'' \cite{Wit}. Nevertheless, one can also provide a
rationale for an interpretation of the RG in connection with
stochastic quantization, independent of string theory. It arises form
the {\em exact} formulation of the RG (including every coupling) which
gives rise to a functional Fokker-Planck equation.

So we begin by describing the exact RG and describing its functional
equation as a Fokker-Planck equation. Then we restrict ourselves to
the usual RG in a finite space of couplings and examine when it can be
considered a gradient flow. In this regard, one must take into account
the freedom in the choice of metric as well as the freedom in the
choice of coordinates. Next, assuming a gradient RG flow, we make the
connection with SUSY QM. Finally, we review Witten's reinterpretation
of Morse theory as SUSY QM and show some applications of Morse
theory to simple examples of RG flow.

\section{The exact renormalization group}
\label{c-grain}

When one says that one is interested in defining the theory at the
scale $L$, one is, first of all, redefining the field $\phi$ to that
scale, by means of an averaging with a suitable kernel:
\be
\phi_L({\bm r}) = \int K_L({\bm r} - {\bm x})\,\phi({\bm x}).
\ee
This is called ``coarse graining''. Customary kernels are the Gaussian
kernel $K_L({\bm x} - {\bm y}) = \exp(-\pi| {\bm x} - {\bm y}|^2/L^2)$
or the ``top-hat'' kernel $K_L({\bm x} - {\bm y}) = 1- \theta(| {\bm
x} - {\bm y}|^2/L^2 -1)$ (where $\theta$ is the step function).  The
first one belongs to the type of ``smooth kernels'', that is, which
are regular functions, whereas the second one does not (for further
explanation, see Ref.~\cite{Ba-Th}). 
We just demand that the kernel has an inverse. In Fourier space,
the coarse graining convolution adopts a simple multiplicative form,
\be
\phi_L({\bm k}) =   K_L({\bm k}) \phi({\bm k}).
\ee
Hence, an inverse exists if $K_L({\bm k})$ has no zeros.

Let us now examine the simple case of the evolution of a Gaussian
probability distribution under a change of $L$. The most general
Gaussian probability distribution can be written as 
\be \P[\phi] =
\exp\{-{1\over 2}\,{\phi}\cdot G^{-1}\cdot {\phi}\} =
\exp\{-{1\over 2}\,{\phi_L\over K_L}\cdot G^{-1}\cdot {\phi_L\over
K_L}\}, 
\ee 
where $G(|{\bm x} - {\bm y}|)$ is the covariance function (the free
propagator in QFT) and we use condensed notation, valid in ``real'' or
Fourier space.  This evolution can be considered trivial: the
coarse-grained field has a variance depressed in the high wavenumbers
$G_L({\bm k}) = K_L({\bm k})^2\, G({\bm k})$.

The evolution of the non-Gaussian part of the probability
distribution with $L$ is more interesting and, not surprisingly,
the calculation leading to it is rather involved: it is the
general form of the Wilson or exact RG. The exact
formulation of the RG was proposed by Wilson \cite{WilKog} and it
has been afterwards the subject of profound studies. We refer the
interested reader to the literature \cite{Ba-Th,Morris}. We are
mainly interested here in the fact that the equation for the
evolution of the non-Gaussian part of the probability distribution
can be written as a linear functional partial differential
equation \cite{WilKog,Morris,I3}: 
\be 
\frac{\p}{\p L} e^{-V_L} = -\frac{1}{2} \frac{\p
G_L}{\p L}\,\frac{\d^2}{\d \phi_L^2} e^{-V_L}, 
\ee 
where $V_L$ is the scale dependent effective potential.
This equation is the simplest form of a functional Fokker-Planck
equation, namely, a functional heat or diffusion equation (a general
functional Fokker-Planck equation including a term with a first
functional derivative results if the Gaussian part is included
\cite{WilKog}).

The essence of coarse graining as we have introduced it is that it
removes the small-scale information in a sort of diffusion process
governed by the usual equations of stochastic dynamics. In particular,
the non-Gaussian part of the probability distribution tends to vanish
in the process, whereas the Gaussian part tends to a fixed form with
only low-${\bm k}$ wavenumbers (in the limit $L \ra
\infty$, only the constant field ${\bm k}= 0$).

The mentioned RG-induced stochastic process takes place in the space
of field configurations and therefore the Fokker-Planck equation is
satisfied by the probability distribution as a function of the field
configuration. We can consider this probability distribution
parametrized by an infinite set of coupling constants in the usual
way. The RG-induced evolution in the space of coupling constants is
deterministic, in principle.  However, if we take into account that
operational forms of the RG can only consider a finite number of
couplings and, therefore, need to truncate the whole space in some
way, we may appreciate that the consequent loss of information must
somehow be added to the inherent loss of information pertaining to
small scales. In fact, both types of information are intertwined,
since the removal of small-scale degrees of freedom leads to the
removal of their couplings.  We conclude that a stochastic process in
the space of couplings follows from the very nature of the
implementation of the RG. We will take advantage of this picture in
the sequel.

\section{The RG as a gradient flow}
\label{gradient}

Here we leave the exact RG and we consider the classical formulation
of the RG as a system of autonomous first order ordinary differential
equations (ODE) for a {\em finite} set of couplings $g_i$ (possibly,
only one): 
\be {d g_i\over d\tau} = \beta_i(g)\,, 
\ee 
where we use a nondimensional RG parameter $\tau$, equivalent to the
logarithm of the normalized relevant scale (e.g., to the logarithm of
the coarse-graining scale $\tau = \log (L/L_0)$). Furthermore, we
consider the situation close to a fixed point $g_i^*$:
\be \beta_i(g) =
\beta_i(g^*) + \left.{\p \beta_i \over \p g_j}\right|_{g^*}\!\! (g_j -
g^*_j) + \cdots, 
\ee
with $\beta_i(g^*) = 0$, so the behaviour of the RG is given by the
linear terms, namely, by the matrix $\Delta_{ij} := \left.{\p \beta_i
\over \p g_j}\right|_{g^*}.$ This matrix is called the dimension
matrix because, when it is diagonalizable ($\Delta_{ij} \ra \Delta_{i}
\delta_{ij}$), the eigenvalues $\Delta_{i}$ give the simple solution
$g_i = g_i(0)\, e^{\Delta_{i}\tau} = g_i(L_0)\, (L/L_0)^{\Delta_i}$
(taking $g^*=0$, for simplicity), so they are proper dimensions.  We
expect the dimensions to be {\em real} positive or negative numbers,
not necessarily integers.  So the generic fixed point is {\em
hyperbolic} (a saddle point).

We may look for general conditions implying that a fixed point has
real dimensions. Obviously, if the dimension matrix is symmetric it
can be diagonalized with real eigenvalues. This is a necessary and
sufficient condition, but without further meaning. A sufficient
condition is that the matrix of derivatives of the beta function is
symmetric in a whole neighbourhood of $g^*$, namely, ${\p \beta_i
\over \p g_j} = {\p \beta_j \over \p g_i}.$ It means that the curl of
$\beta_i$ vanishes, so that it is the gradient of some function
$V(g)$: $\beta_i = {\p V \over \p g_i}$; this is called a gradient
flow \cite{WallZia}. It follows that ${d V\over d\tau} = \beta^2 \geq
0$, that is, $V$ is a monotonic (Lyapunov) function of the system of
ODE.

Geometrically speaking, the gradient flow is orthogonal to the
(hyper)surfaces of constant $V$. This orthogonality depends on a
metric, which has been taken to be euclidean by default. In fact,
covariance demands that the gradient flow condition be written as
$\beta^i = G^{ij}{\p_j V}$, where the metric can be arbitrary. 
So if we are given a flow the question of whether it is a gradient 
flow or not is somewhat ambiguous and can be interpreted as the
question of whether there can be found a metric that makes it a
gradient flow.
Now, it is easy to convince oneself that if we allow for an
arbitrary metric we can always conclude that a flow is gradient
near a fixed point if and only if the dimensions are real. 
Of course, we should then ask for a {\em
natural} metric and that it be {\em globally} defined. 

Therefore, we can express the gradient flow condition in an 
intrinsic form: since the RG flow is given by a vector field 
$\beta$ on a manifold, it is a gradient flow if for some metric $G$ 
in the manifold the one-form $\theta = G(\beta)$ is exact, namely, 
$\theta = dV.$ Furthermore, we expect to have a natural metric.
Indeed, there is a natural metric in the space
of coupling constants when these are considered as statistical
parameters: the Fisher metric of estimation theory \cite{Ama}. The
quest for RG gradient flows with this metric has already had
partial success \cite{Dolan,Bro}. Furthermore, the relation of Fisher
metric with entropy (or information) constitutes the basis for an
extension of the time-irreversibility $H$-theorems to
irreversibility under scale transformations (that is, under the RG)
\cite{IOCon}. We must also mention that in two dimensions there is
another candidate for a natural metric, namely, Zamolodchikov's
metric \cite{Zamo}. Intriguingly, the Fisher metric (valid in any
dimension) and Zamolodchikov's metric adopt somewhat similar
expressions \cite{I1}.

\subsection{Freedom in the choice of coordinates. Scaling fields}

We have mentioned that the RG must act covariantly in the space of
couplings; in other words, we are free to choose coordinates in
this space, redefining the couplings. This large freedom implies
in particular that we can always make the RG a gradient flow by
linearizing the $\beta$ functions: the corresponding coordinates
are called scaling fields (scaling is homogeneous in these
coordinates) \cite{WallZia}. The possibility of linearizing a flow is
in fact a general result of the theory of ODE, in which it is
called Poincar{\'e} theorem \cite{Arnold2}. 
In quantum field theory the scaling
fields are to be identified with the bare couplings. The bare
couplings indeed scale with their naive dimensions whereas the
behavior of renormalized couplings under a change of scale is
given by the beta functions, including ``anomalous" dimensions.

The simplest example is perhaps the RG for the theory
$\l\,\phi^4$ (in dimension $D<4$). The one-loop RG equation can
be written as \cite{ZJ}
\begin{equation}
{{d \l\over d\tau} = -\mu {d \l\over d\mu} = \l - \l^2},
\end{equation}
after linearly redefining both $\tau$ and the coupling $\l$ to
make numerical coefficients equal to one. These redefinitions
place the fixed points at $\l = 0,\, 1$. The solution of this
equation with the condition $\l(1)= \l_0 \in (0,1)$ is
\begin{equation}
{\l = \frac{\l_0}{\mu + \l_0 (1 - \mu)}\,.}
\end{equation}
It gives the flow between the two fixed points. In the UV limit
$\mu \ra \infty$, $\l \ra 0$ but $\mu\l$ stays finite. We can
define the scaling coupling {$$\l_b = \lim_{\mu \ra \infty}(\mu\l)
= \frac{\l_0}{1 - \l_0}\,.$$} Indeed,
\begin{equation}\label{project}
{\tilde{\l} = \frac{\l}{1 - \l}}
\end{equation}
is the coordinate transformation that linearizes the RG:
\begin{equation}
{\mu {d \tilde{\l}\over d\mu} = -\tilde{\l}\: \Ra \:\tilde{\l} =
\frac{\l_b}{\mu}}\,.
\end{equation}
Note that in the scaling coordinate the IR fixed point is located
at $\tilde{\l} \ra \infty$.

Let us remark that transformation (\ref{project}) is projective
(an $RP^1$ mapping). This is no coincidence: in general, one-loop
RG equations implement projective transformations of the
couplings and the real projective space is the natural
compactification of the space of couplings \cite{IOCon}. While in
scaling coordinates the nontrivial fixed points (that is, other
than the one at the origin) are located at infinity and there is a
trivial quadratic potential for the RG flow, in coordinates that
cover the nontrivial fixed points the potential becomes
nontrivial. Furthermore, projective space is homogeneous with {
its natural metric}. Hence, this metric is also the natural
metric for covariant gradient RG flow.

      \section{Stochastic RG and SUSY in the space of couplings}

Let us assume that the RG is a gradient flow in a finite
dimensional space of couplings, and that the state of the system
is represented by a probability distribution on this space, as
remarked in section \ref{c-grain}. Hence, we can derive interesting
consequences.

The implementation of the RG leads to loss of information on the
couplings, so that the exact state given by the infinite coupling
constants becomes a probability distribution $P(g^i)$ over a finite
set of couplings (as remarked at the end of section
\ref{c-grain}) and the RG evolution of these couplings can be
represented by adding stochastic components to the
$\beta$-functions. This makes the RG equations into Langevin
equations:
\begin{equation}
{d g^i\over d\tau} = \beta^i(g) + \eta^i\, , \quad \langle
\eta^i(\tau) \eta^j(\tau') \rangle = G^{ij}(g) \d(\tau - \tau').
\end{equation}
The noise is assumed to be white and we have introduced the metric
for covariance. $P(g^i)$ satisfies a Fokker-Planck equation,
associated with the preceding Langevin equations \cite{ZJ}.
Assuming that $\beta_i(g) = \p_iV(g)$ (corresponding to a purely
dissipative Langevin equation), the Fokker-Planck equation can be
expressed as a Schr\"odinger equation in imaginary time with a
{\em hermitian} hamiltonian.

Let us first express the Fokker-Planck equation as
\begin{equation}
{\p \P(g,\t)\over \p\tau} =
-H_{FP} \P(g,\t) = \frac{1}{2} \nabla^2 \P(g,\t) -
\nabla_i\left(\beta^i\,\P(g,\t)\right).
\end{equation}
Then, the hermitian hamiltonian is
\begin{equation}
\tilde H_{FP} = e^{-V} \,H_{FP}\; e^{V}
= -\frac{1}{2}\left(\nabla^2 - (\nabla V)^2 - \nabla^2 V\right) =
\frac{1}{2} A_i^{\dagger}A_i\,,
\label{HFP}
\end{equation}
with ${\bm A} = \nabla - \nabla V$. This hamiltonian operates on
states $|g,\t\rangle = e^{-V} \P(g,\t)$, the equilibrium state
being $\langle g,\t| 0 \rangle = e^{V}$, such that ${\bm A}|
0 \rangle = 0$. ``Excited states'' are produced by the action of ${\bm
A}^{\dagger}$.

      \subsection{Connection with SUSY QM}

Let us now summarize Das, Mandal and Wadia's procedure to represent
the probability distribution $P(g,\tau)$ as SUSY QM, following
Parisi-Sourlas's \cite{ParSour} and Witten's \cite{Wit} methods:
\begin{enumerate}

\item {Introduce fermionic coordinates $\psi^i(\tau)$ and
$\bar\psi^i(\tau)$ (Grassmann variables), such that $(\psi^i)^2 =
[\bar\psi^i]^2 = 0, \quad \{\psi^i,\bar\psi^j\} = G^{ij}.$}

\item Introduce supercharges {$Q = \sum_i \psi^i (\nabla_i +
\beta_i)$} and $\bar Q$. Note that $Q^2 = (\bar Q)^2 = 0$ if and only if
{$\beta_i = \p_i V$}, that is, if the RG is a {gradient flow}.

\item {Complete the SUSY algebra with the supersymmetry
hamiltonian}
\begin{equation}
H = \frac{1}{2}(Q\bar Q +\bar Q Q) = 
\frac{1}{2} \left(-\nabla^2 + G^{ij} \p_iV \p_jV +
\nabla_{i}\nabla_jV\, [\psi^i,\bar\psi^j]\right). \label{H}
\end{equation}
This is just the SUSY generalization of the hamiltonian
(\ref{HFP}).
\end{enumerate}

The euclidean action corresponding to the preceding hamiltonian is 
\cite{Wit,DasManWa}
\begin{equation}
\S = \frac{1}{2}\int d\t
\left[G_{ij}\left(\frac{dg^i}{d\t} \frac{dg^j}{d\t}
+ \p^iV \p^jV \right) -
G_{ij} \bar\psi^i \frac{dg^k}{d\t} \nabla_{k}\psi^j +
\frac{1}{4} R_{ijkl} \bar\psi^i \psi^k \bar\psi^j \psi^l
+ \nabla_{i}\nabla_jV\,\bar\psi^i\psi^j \right],
\label{fermS}
\end{equation}
in which appear the Riemann curvature tensor, etc. 
This is the action of
a {\em one-dimensional} $N=2$ supersymmetric nonlinear
$\sigma$-model.
It can be written 
in terms of the supercoordinate
$\phi(\t) = g(\t) + i\th \psi(\t) - i\bar\psi(\t) \bar\th + \bar\th
\th \,\nabla V$:
\begin{equation}
\S = \int d\t\, d\bar\th\, d\th
\left(\frac{1}{2}D\phi \cdot\overline{D\phi} - V(\phi) \right),
\end{equation}
where $D = \p_\th - \bar\th \p_\t$, and $V$ is the superpotential.

The bosonic part of the action is
\begin{equation}
\S = \frac{1}{2}\int d\t \;G_{ij}\left(\frac{dg^i}{d\t} \frac{dg^j}{d\t}
+ \p^iV \p^jV \right).
\label{boseS}
\end{equation}
The analysis of the minima of this action has important consequences
for the topology of the space of couplings.
Indeed, it is easy to see \cite{Wit} that the minima occur for
\begin{equation}
\frac{dg^i}{d\t} \pm \p^iV = 0\,, \label{clas}
\end{equation}
which defines the gradient flow (in either direction).

\section{Morse theory and topology of the space of couplings}

Morse theory \cite{Milnor}
extracts topological information on a manifold from
the knowledge of the critical points of some function on the
manifold. Conversely, if the topology of the manifold is known,
one can use it to deduce the existence and properties of the
critical points of a function. Morse theory is generally applied
to finite dimensional manifolds but it has also been used in some
infinite dimensional spaces \cite{Vafa}.

The question in our case is what space we should consider. The
basic infinite dimensional space seems to be the projective space
of probability distributions, its projective character coming from
the normalization of the probability distribution \cite{Bro}.
However, the RG flow that we are considering operates in a finite
dimensional subspace. As we mentioned in section \ref{gradient},
the natural geometry seems to correspond to a real projective
space $RP^n$ (see also \cite{Bro}). The topology of the real
projective space is well known, so we can deduce the properties of
the critical points of any potential defined on that space. In
order to see how to proceed, let us review Witten's
reinterpretation of Morse theory as SUSY QM \cite{Wit}.

Witten considers the fermion coordinates $\psi^i$ and $(\psi^*)^i$
as operators on the exterior algebra acting by interior and
exterior multiplication, respectively. The basic objects in the
algebraic topological theory by means of de Rham cohomology are
the exterior derivative $d$, its adjoint $d^*$, and the Hodge
Laplacian $\Delta = (d + d^*)^2$. The supersymmetry operators $Q$
and $Q^*$ are then interpreted as new exterior derivatives
obtained from $d$ and $d^*$ by conjugation with the exponential of
a function $V$, namely, $Q = e^{-V} d\,e^{V}$ and $Q^* = e^{V}
d^*\,e^{-V}$. Thus the hamiltonian (\ref{H}) is the transform of
the Hodge Laplacian. It is easy to prove that this transformation
is a isomorphism of the exterior algebra, so the algebraic
topological properties are left unchanged by it. Furthermore, if
we consider the classical limit, that is, when the noise
fluctuations are negligible and the classical equations
(\ref{clas}) hold, the isomorphism is still valid, so we deduce
that the topological properties can be extracted from the critical
points of $V$.

Morse theory assumes that the critical points of $V$ are
non-degenerate, that is, the Hessian determinant is nonvanishing at
them. The topological information is encoded in the {\em index} of $V$
at the critical points, which is defined as the number of negative
eigenvalues of the Hessian matrix. In fact, Morse lemma shows that in
a neighborhood of a critical point exist local coordinates such that
the function is a quadratic form (of course, related with the scaling
coordinates of section \ref{gradient}) and, furthermore, that the
coefficients can be made to be $\pm 1$.  Therefore, the only
topological information is in the relative number of negative and
positive coefficients, that is, the index.  One then associates to $V$
and its critical points the Morse polynomial
\begin{equation}
M(V) = \sum_{P_i} t^{n_i},
\end{equation}
where $P_i$ are the critical points and $n_i$ their respective
indices. The topology of the manifold enters via the Poincar\'e
polynomial
\begin{equation}
P = \sum_{i=0}^n b_i t^{i},
\end{equation}
where $b_i = \dim H^i$ are the Betti numbers, defined as the
dimensions of the cohomology groups. The fundamental result of Morse
theory (Morse inequalities) is that $M(V) \geq P$ and, moreover,
\begin{equation}
M(V)- P = (1+t) Q,
\label{MorseP}
\end{equation}
where $Q$ is a polynomial with positive coefficients. A function
$V$ for which $M(V)= P$ is called a perfect Morse function. For
every compact finite dimensional manifold one can find a perfect
Morse function.

An interesting consequence of Eq.~(\ref{MorseP}) occurs for $t=-1$, namely, 
\begin{equation}
M(V)(-1) =  P(-1) = \sum_{i=0}^n (-1)^{i} b_i = \chi \,,
\end{equation}
that is, the Euler-Poincar\'e characteristic. Therefore, the 
Poincar\'e-Hopf index theorem on the zeros of a vector field
\cite{Arnold1} of gradient type 
is a particular case of Eq.~(\ref{MorseP}) (note that $V$ needs not 
be a perfect Morse function).

\section{RG gradient flows with one or two couplings}
 
An elementary application is the theory $\l\,\phi^4$ considered
in section \ref{gradient}. The potential $V$ in the scaling
coordinate seems to be just $V=\tilde\l^2/2$ but we must account
for the metric of $RP^1 = S^1/Z_2 \simeq S^1$ 
(in general, $RP^n = S^n/Z_2$, where the $Z_2$ factor is to identify 
antipodal points). 
The metric in this coordinate is $ds^2 = d\tilde\l^2/(1+\tilde\l^2)^2$. 
So the correct potential is
\begin{equation}
V = \frac{\tilde\l^2}{2\,(1 + \tilde\l^2)}\;,
\end{equation}
which coincides with $\tilde\l^2/2$ when $\tilde\l \ll 1$ and
has finite limit when $\tilde\l \ra \infty$. Note that the
critical points of $V$ are $\tilde\l = 0, \;\infty$, that is, both
RG fixed points. The Morse polynomial is simply $M(V) = 1 + t$.
Naturally, the Poincar\'e polynomial of $RP^1$ is also $P = 1 + t$
so $V$ is a perfect Morse function.

The function $\beta(\l)$ of the $\l\,\phi^4$ theory at more than
one loop order is a higher degree polynomial, so it may have more than
two fixed points and then corresponds to a potential $V$ with several
extrema. If this happens, $M(V)$ also becomes a higher degree
polynomial, so $V$ is no more a perfect Morse function. At any rate, it
is questionable the validity of perturbation theory for finding the
additional nontrivial fixed points, being only important the first one
(at any loop order).

A somewhat less elementary application is the theory for tricritical
behavior $r\, \phi^2 + \l\,\phi^4 + g\,\phi^6$ (in dimension $3 \leq
D<4$) \cite{IOCon}.  The RG equations for the relevant bare couplings 
(the scaling coordinates) are just
\begin{eqnarray}
\frac{d\tilde r}{d\t} = \varphi\,\tilde r\,,
\label{RGe1}\\
\frac{d\tilde\l}{d\t} = \tilde\l\,.
\label{RGeq1}
\end{eqnarray}
The trajectories are given by $\tilde r \propto \tilde\l^\varphi$.
The crossover exponent $\varphi > 1$ can be taken to be 2 (the
mean-field value for $D=3$) without loss of generality.
Under the projective change of coordinates
\begin{eqnarray}
\label{project2-1}
\tilde{r} = \frac{r}{1 - r - \l}\;,\\
\label{project2-2}
\tilde{\l} = \frac{\l}{1 - r - \l}\;,
\end{eqnarray}
the RG equations become
\begin{eqnarray}
\frac{d r}{d\t} = r (2(1 - r) - \l)\,,
\label{RGe2}\\
\frac{d\l}{d\t} = \l (1 - \l - 2 r)\,.
\label{RGeq2}
\end{eqnarray}
Similar equations were derived in Ref.~\cite{NiChaStan} from the
Wegner-Houghton RG. 

The advantage of coordinates (\ref{project2-1},\ref{project2-2}) and
the preceding RG equations is that the fixed points are at finite
positions, namely, the tricritical point is at $r=\l=0$, the critical
point is at $r=0,\,\l=1$, and the high-temperature Gaussian point is
at $r=1,\,\l=0$. However, the scaling coordinates are simpler for
deriving the RG potential. To do this, we must consider the $RP^2$ (or
$S^2$) metric, namely,
$$ds^2 = \frac{1}{(1+\tilde r^2+\tilde\l^2)^2} 
\left((1+\tilde\l^2) d\tilde r^2 + (1+\tilde r^2) d\tilde\l^2
-2 \,\tilde r\, \tilde\l\, d\tilde r \,d\tilde\l \right).$$
We obtain
\begin{equation}
V = \frac{\tilde r^2 + \tilde\l^2/2}{1 + \tilde r^2 +\tilde\l^2}
= \frac{r^2 + \l^2/2}{1 + 2 r^2 + 2\l^2 - 2\,\l(1-r) -2 r}
\label{pot2}
\;.
\end{equation}
Since we have a minimum, a saddle point and a maximum, the Morse
polynomial is $M(V) = 1 + t + t^2$.  The Poincar\'e polynomial of
$RP^2$ is also $P = 1 + t + t^2$ so $V$ is a perfect Morse function.

We remark that this flow on $RP^2$ has three invariant subspaces
$RP^1$, corresponding to the tricritical-critical crossover, the
tricritical-Gaussian crossover, and the critical-Gaussian crossover.
The tricritical-critical crossover occurs for $\tilde r = r =0$.  The
corresponding RG equations (\ref{RGeq1}), (\ref{RGeq2}), and potential
(\ref{pot2}) coincide with the ones of the $\l\,\phi^4$ theory.

It is pertinent here to relate the preceding remark with the previous
remark about additional nontrivial fixed points in $\l\,\phi^4$
theory. We see that to have more meaningful nontrivial fixed points in
this theory we must introduce an additional coupling (making the
coupling space two-dimensional). This is not the case for
two-dimensional flows, that is, field theories with two couplings can
have a fairly complicated fixed point distribution, corresponding to
more complicated topologies. For example, the next more complex case
than $RP^2$ is the torus or the Klein bottle, both corresponding to a
potential with two nodes and two saddle points.

As an example of a theory with four critical points, consider a
two-component theory, namely, $\l_1\phi_1^4 + \l_2\phi_2^4$, in which 
the two fields do not necessarily have the same dimension. The
RG equations for scaling couplings are 
\begin{eqnarray}
\frac{d\tilde \l_1}{d\t} = \varphi\,\tilde \l_1\,,\\
\frac{d\tilde \l_2}{d\t} = \tilde \l_2\,,
\end{eqnarray}
equivalent to Eqs.\ (\ref{RGe1}, \ref{RGeq1}).  However, it may happen
that the RG equations for the renormalized couplings do not admit
crossed terms; that is, in the equations corresponding to Eqs.\
(\ref{RGe2}, \ref{RGeq2}) the crossed terms are missing. This
indicates that the relation between scaling and renormalized couplings
consists of {\em independent} projective transformations for $\l_1$
and $\l_2$. Then there are three nontrivial fixed points, namely,
$(\l_1=0,\,\l_2=1)$, $(\l_1=1,\,\l_2=0)$, and $(\l_1=1,\,\l_2=1)$; the
first and the second are saddle points while the fourth is a node. The
corresponding compactified coupling space is the direct product $RP^1
\times RP^1$, namely, the torus.  The corresponding Poincar\'e
polynomial is $P = (1+t)^2 = 1 + 2t + t^2$ and the Morse polynomial is
also $1 + 2t + t^2$, so $$V= \frac{1}{2}
\left(\frac{\varphi\,\l_1^2}{(1 - \l_1)^2+
\l_1^2} + \frac{\l_2^2}{(1 - \l_2)^2 + \l_2^2}\right)$$ is a
perfect Morse function.

      \section{Discussion}

We have seen that the exact formulation of the RG provides us with an
instrument to analyze the evolution of the infinite number of
couplings of a field theory. However, this infinite-dimensional
coupling space is too complex to study, except maybe in the case of
two-dimensional field theories \cite{Vafa}, and implementations of the
exact RG must truncate it to a finite-dimensional space of couplings
\cite{Morris}. The infinite number of neglected irrelevant couplings
produce some uncertainty in the values of the preserved couplings, so
it is necessary to add {\em noise} to the RG equations and,
therefore, to substitute a definite location on coupling space by a
probability distribution (in the context of string theory
\cite{DasManWa}, this is equivalent to second quantization).

The preceding substitution of a definite location in coupling space by
a probability distribution has consequences on the issue of RG
irreversibility. As in the classical statistical theory of time
irreversibility associated with the neglect of microscopic degrees of
freedom in a macroscopic description, we have that the probability
distribution in coupling space evolves irreversibly, as the
corresponding Langevin or Fokker-Planck equations attest, and that the
RG potential plays the role of irreversible function, which in general
has entropic nature \cite{IOCon,I1}.

The introduction of a stochastic formulation for the RG may bring some
complications but it also allows one to connect with the
supersymmetric formulation of stochastic quantization. In particular,
if the RG $\beta$-function is the gradient of a potential, the
stochastic RG is equivalent to SUSY QM in the finite space of
couplings and, hence, one can study the topology of this space by
means of Morse theory with the potential and, viceversa, one can
deduce the types of fixed points from the topology.

The simplest candidate for compactification of $n$-dimensional
coupling space is $RP^n$, whose topology is well-known. Hence, it is
possible to deduce general patterns for RG flows. The study of the one
and two-dimensional cases shows how simple field theory RG's adapt to
$RP^n$ for $n=1,2$.  Presumably, a generic RG flow will have the
topology of the gradient flow given by a perfect Morse function on
$RP^n$.  More complex RG flows may correspond to subspaces of it, like
the two-dimensional torus already described.  This case is
particularly interesting, because the RG equations for scaling
couplings are indistinguishable from the respective equations leading
to $RP^2$.  This shows the importance of the topology, that is, how
different compactifications lead to {\em globally} different flows.
The relation between perturbative renormalization and the global
character of RG flows is a subject that deserves further study.

\subsection*{Acknowledgements}

My work is supported by a Ram\'on y Cajal contract and by grant
BFM2002-01014, both of Ministerio de Ciencia y Tecnolog\'{\i}a.  I
thank Luis J. Boya for a conversation on SUSY QM and Oscar
Garc\'{\i}a-Prada and Juan Mateos-Guilarte for comments on the
manuscript.

\end{document}